\documentclass[aps,twocolumn,amsmath,amssymb,preprintnumbers,longbibliography]{revtex4-1}
\usepackage{amsmath} \usepackage{amsfonts} \usepackage{amssymb}
\usepackage{titlesec}
\usepackage{bbm}
\usepackage{epsfig}
\usepackage{graphics}
\usepackage{graphicx}
\newcommand{\be}{\begin{equation}}
\newcommand{\ee}{\end{equation}}
\newcommand{\ba}{\begin{eqnarray}}
\newcommand{\ea}{\end{eqnarray}}
\newcommand{\nn}{\nonumber}

\newcommand{\m}{_{\mu\nu\rho\sigma}}
\newcommand{\ti}{t_{\rm in}}
\newcommand{\ri}{_{\rm in}}
\newcommand{\f}{^{\frac{4}{2-A}}}
\newcommand{\nr}{_{nr}}

\titleformat{\subsection}[block]{\normalfont\bfseries}{\thesubsection.}{1ex}{}
\titlespacing{\subsection}{0pt}{10pt}{1pt}[0pt]
\titleformat*{\section}{\large\bfseries}
\renewcommand{\thesubsection}{\arabic{subsection}}

\begin{document}

\title[ ]{Eternal Universe}

\author{C. Wetterich}
\affiliation{Institut  f\"ur Theoretische Physik\\
Universit\"at Heidelberg\\
Philosophenweg 16, D-69120 Heidelberg}

\begin{abstract}
We discuss cosmological models for an eternal universe. Physical observables show no singularity from the infinite past to the infinite future. While the universe is evolving, there is no beginning and no end - the universe exists forever. The early state of inflation is described in two different, but equivalent pictures. In the freeze frame the universe emerges from an almost static state with flat geometry. After entropy production it shrinks and ``thaws'' slowly from a ``freeze state'' with extremely low temperature. The field transformation to the second ``big bang picture'' (Einstein frame) is singular. This ``field singularity'' is responsible for an apparent singularity of the big bang. Furthermore, we argue that past-incomplete geodesics do not necessarily indicate a singularity or beginning of the universe. Proper time ceases to be a useful concept for physical time if particles become massless. We propose to define physical time by counting the number of zeros of a component of the wave 
function. This counting is independent of the choice of coordinates and frames, and applies to massive and massless particles alike.  

\end{abstract}

\maketitle
\section{Introduction}
\label{Introduction}

Can the universe exist forever, without beginning and end? Since the failure of steady state cosmologies and the general acceptance of the big bang it is widely believed that the universe must have had some type of ``beginning''. The Friedmann-Lema\^{i}tre cosmological solution becomes singular as the big bang is approached. It can therefore not be extended to an infinite past. Assuming the strong energy condition Penrose and Hawking have shown the presence of a past singularity or geodesic incompleteness for rather arbitrary cosmological solutions \cite{Pen,Haw}. With the advent of inflation the strong energy condition has been abandoned. Still, with the rather mild assumption that the universe is expanding in the average (more precisely, that the average Hubble parameter is positive) it has been established that geodesics cannot be complete towards the past \cite{BGV,AV}. From this observation the conclusion was drawn that the universe becomes singular in the finite past, or at least cosmology becomes 
incomplete,
 necessitating a ``beginning''. For a a wide class of inflationary models or alternative ``pre big bang models'' an extension to the infinite past seems unfeasible. 

Recently, simple models have been proposed \cite{CWUE,CWSF} for which no past singularity occurs. These cosmologies can be extended to the infinite past. In terms of only a few parameters these models can describe all present observations, including inflation, an end of inflation, radiation - and matter-domination and the present transition to a new dark energy dominated period. They are thus fully consistent and constitute counter examples to the view that the universe must have had a beginning.

The evolution of the universe is typically very slow in these models - the characteristic time scale is never much shorter than the present inverse Hubble parameter $\sim 10^{10}$yr. The geometry approaches flat space in the infinite past. All geometrical invariants built from the curvature tensor and its covariant derivatives, contracted with the inverse metric, vanish for the infinite past, $t\to-\infty$. In this ``freeze picture'' it seems rather obvious that no singularity is encountered, with a cosmological solution extending to the infinite past. Nevertheless, the same models can be mapped by a conformal transformation (Weyl scaling) to an equivalent ``big bang picture''. In this Einstein frame the primordial cosmology is of a standard inflationary type. Field relativity \cite{CWUE,CWQ1} states that the two pictures are fully equivalent. The absence or presence of physical singularities should be the same 
in both pictures. The big bang picture has a geometry with geodesics that become incomplete in the past. If the presence of incomplete geodesics would really indicate  a physical singularity it would be hard to understand how the freeze picture could be free of singularities.

In this note we address the connections between incomplete geodesics, curvature singularities, and the possible existence of physical singularities in the light of transformations between different frames. This will shed new light on the role of ``singularity theorems''. The discussion will lead to four central findings:
\begin{itemize}
\item [(i)]
Field transformations, as the conformal transformation between different frames, can be singular. A detected singularity in some frame may therefore arise from a singularity in the field transformation, while in some other frame everything is regular. Such ``field singularities'' do not reflect a physical singularity, in analogy to ``coordinate singularities'' arising from the choice of a particular coordinate system. (They are singularities in ``field-coordinates''.) The absence of physical singularities is guaranteed if one frame exists where all relevant physical observables are found to be regular. 
\item [(ii)]
Cosmological solutions can have attractor properties. As a consequence, after an evolution over a certain time interval only a restricted range of field values and their derivatives will be found at some given time $t_0$. Inversely, if one tries to extrapolate backwards, with ``initial conditions'' at $t_0$ outside this allowed range, one typically encounters a singularity. Even for a regular universe the most general solutions with arbitrary ``initial conditions'' at $t_0$ will not remain regular towards the infinite past. In this case the presence of singular solutions neighboring a regular solution should not be misinterpreted as a sign that a ``beginning'' of the universe is needed. For example, an attractive regular isotropic solution may be surrounded by anisotropic solutions that become singular in the past. 
\item [(iii)]
Physical time must not only be coordinate independent but also frame independent. Frame independent quantities are dimensionless, as proper time multiplied by the particle mass, evaluated on the trajectory of a massive particle. Proper time by itself is changed by field transformations. Even dimensionless proper time is no longer a useful physical clock if the ratio momentum/mass diverges. In this case a particle behaves like a photon. A reasonable coordinate and frame independent physical time may be defined by counting the number of oscillations of a wave function. 
\item [(iv)] The presence of timelike geodesics that are incomplete towards the past does not necessarily indicate a singularity or incompleteness of cosmology.  Particles behave as massive particles only for a finite ratio momentum/mass, and only in this case proper time is a useful measure of time. For finite momentum/mass in the infinite past the allowed velocities $u(t_0)$ at some finite time point $t_0$ are restricted.  Past-incomplete geodesics can be precisely those with $u(t_0)$ outside the allowed range. In this case particles behave like photons in the infinite past and proper time ceases to be a useful measure of physical time. 
\end{itemize}

We start by specifying our criteria for an eternal cosmology that is free of singularities from the infinite past, $t\to-\infty$, to the infinite future, $t\to\infty$: (i) The cosmological solution should be regular for all $t$. (ii) For a suitable definition of physical time the time-distance to the infinite past and infinite future should both be infinite. (iii) For massive particles and in suitable units the proper time elapsed from some given time $t_0$ to the infinite future should be infinite. (iv) Also the proper time from the infinite past to $t_0$ should be infinite if momentum/mass remains finite. (v) Furthermore, we require that no trajectory of a massive or massless particle encounters a singularity in the whole range between the infinite past and future. 

A few comments on these criteria are in order: For momentum/mass $\to\infty$ particles behave as photons and proper time becomes unsuitable. The condition for the use of proper time for measurements of physical time may be weakened by requiring only finite momentum and a finite suitable time averaged ratio momentum/mass, such that particles do not behave as photons for most of their 
history. Obviously, the ``eternity'' of the universe has to be defined in a coordinate-independent concept as proper time or ``oscillation time''. We can always choose a time coordinate $t$ that covers an infinite range from $-\infty$ to $+\infty$, even for a cosmology with a physical singularity.

Our general strategy is rather simple. For a given model we first consider the freeze frame where it is rather  easy to get convinced that observables remain regular from the infinite past to the infinite future. The singular map to the Einstein frame is then used to understand the singularities of the big bang as an inappropriate choice of time or geometry. These singularities appear to be field singularities, while physical observables remain regular.

We demonstrate our points with two specific models of gravity coupled to a scalar field. This field is responsible for both inflation and late dark energy. A crossover between two fixed points is responsible for the transition from the inflationary primordial cosmology to ``late cosmology''. In sect. \ref{Crossover model with flat space in the infinite past} we present our first model which admits asymptotic solutions where the infinite past corresponds to Minkowski space with constant scale factor. For $t\to-\infty$ the Hubble parameter vanishes $\sim(-t)^{-3}$ while particle masses go to zero with a different inverse power of $-t$. The geometry is obviously regular and geodesically complete. 

In sect. \ref{Focus property of primordial cosmology} we show that this family of asymptotic solutions is an attractor for increasing time, to which neighboring isotropic and homogeneous cosmologies converge. As a consequence, the general solution with arbitrary integration constants fixed at some finite $t_0$ cannot 
be 
continued to the infinite past - this is only possible for 
the family of attractor solutions. In sect. \ref{Physical time} we address possible definitions of physical time in this cosmology. We find that proper time evaluated on the trajectories of massive particles is not suitable for this purpose. All masses vanish in the infinite past such that particles with nonzero momentum behave as photons. We propose to use instead the counting of the oscillations of the wave function which works both for massive and massless particles. For this coordinate- and frame-invariant ``physical time'' both the infinite past and future are at infinite distance.

The model is mapped to the Einstein frame in sect. \ref{Singularities in the Einstein frame}, where our solutions describe power-law inflation. The conformal transformation of the metric becomes singular in the infinite past, which is the root of the apparent singularities in the big bang picture. While particle trajectories are mapped to particle trajectories, this does not hold for geodesics. Also proper time is not invariant under a change of frame, while the counting of oscillations remains the same in all frames. For ``oscillation time'' the geometric singularity remains in the infinite past. In turn, this geometric singularity is due to a particular choice of metric, while for a different choice the geometry remains regular. 

In sect. \ref{Crossover model for de Sitter inflation} we turn to our second model which describes de Sitter inflation in the Einstein frame. In the freeze frame the scale factor vanishes faster than a power and slower than an exponential, with vanishing curvature invariants in the infinite past. For this model singularities in the curvature invariants are absent in both frames, despite the singularity of the conformal transformation. We discuss in detail the interpretation of geodesic incompleteness of de Sitter space. It is linked to the property that particles with finite momentum become photon-like in the infinite past, such that proper time is no longer suitable for a definition of physical time. 

We demonstrate in sect. \ref{Eternal Universe and inflation} that the class of crossover models to which our two models belong are viable candidates for an inflationary epoch of the universe. For this purpose we discuss a whole family of models that interpolate between the two models of sects. \ref{Crossover model with flat space in the infinite past} and \ref{Crossover model for de Sitter inflation}. They lead to realistic scenarios for inflation, typically with large tensor fluctuations. Our models are therefore not only rather simple examples for an eternal universe. They can also be taken as realistic candidates for the description of our observed world. Our conclusions are presented in sect. \ref{Conclusions}.

\section{Crossover model with flat space in the infinite past}
\label{Crossover model with flat space in the infinite past}

Our two models belong to a family of variable gravity models \cite{CWQ1,CWQ2,CWVG,HMSS} where the effective value of the Planck mass (or gravitational constant) depends on a scalar ``cosmon'' field $\chi$. They are specified by the quantum effective action
\be\label{1a}
\Gamma=\int_x\sqrt{g}\left\{-\frac12\chi^2R+\frac12(B-6)\partial^\mu\chi\partial_\mu\chi+V(\chi)\right\}.
\ee
The effective gravitational ``constant'' is always positive - no antigravity occurs. A constant $B>0$ guarantees stability provided $V$ is bounded from below. For the potential we assume a crossover from a behavior $V\sim \chi^{4-A}$ for $\chi\to 0$ to $V=\mu^2\chi^2$ for $\chi\to\infty$, namely
\be\label{2a}
V=\frac{\mu^2\chi^{4-A}}{m^{2-A}+\chi^{2-A}}\quad,\quad 0\leq A\leq 2.
\ee
For the primordial cosmology that extends to $t\to-\infty$ only the behavior for $\chi\to 0$, $V_0\sim\chi^{4-A}$, will be needed. The crossover to a different behavior for $\chi^2\gg m^2$ is required, however, to end the early inflationary epoch, making a transition to realistic radiation - and matter-domination. For $\chi^2\ll m^2$ we assume that the masses of all particles are proportional to $\chi$. 

\subsection{Minkowski space for the infinite past}
Our first model takes $A=B,~A\ll 1$. It has two dimensionless parameters $A$ and $\mu/m$. We will see that for large negative $t$ geometry approaches flat Minkowski space, with Hubble parameter going to zero as
\be\label{2AA}
H=-\frac{h\mu}{(1-\mu t)^3}.
\ee
It seems rather obvious that such a geometry is ``past eternal''. 

For a discussion of inflationary primordial cosmology and the end of inflation we neglect matter and radiation. We have to solve the field equations for the coupled cosmon-gravity system that follow from variation of $\Gamma$. The modification of gravity due to the variable Planck mass induces new features, as a ``driving force'' for the evolution of $\chi$ proportional to the curvature scalar $R$. For a (spatially flat) Robertson-Walker metric with scale factor $a(t),~H=\partial_t\ln a$, and a homogeneous cosmon field $\chi(t)$, the cosmon field equation reads \cite{CWVG}
\be\label{3a}
\ddot{s}+3H\dot s+2\dot s^2=\frac{\mu^2x(A+2x)}{A(1+x)^2},
\ee
with 
\be\label{4a}
s=\ln\frac\chi m\quad , \quad x=\left(\frac\chi m\right)^{2-A}=e^{(2-A)s}.
\ee
Here we have already inserted the curvature scalar $R$ according to the gravitational field equation. The Hubble parameter obeys
\be\label{5a}
H=\sqrt{\frac{\mu^2x}{3(1+x)}+\frac{A\dot s^2}{6}}-\dot s.
\ee

For $t\to-\infty$ the coupled system of equations \eqref{3a}, \eqref{5a} admits a simple approximate solution \cite{CWVG}
\be\label{6a}
\frac\chi m=
\left[\frac{(2-A)}{\sqrt{12-2A}}(1-\mu t)\right]^{-\frac{2}{2-A}}~,~H=0, 
\ee
for which geometry is Minkowski space. This solution becomes exact if $V$ is approximated by $V_0=\mu^2 m^{A-2}\chi^{4-A}$. In the infinite past $\chi$ approaches zero. There exists another exact solution $\chi=0, H=0$. It is unstable, however, with a small deviation $\chi$ increasing according to eq. \eqref{6a}. 

Due to the crossover form of the potential \eqref{2a} the cosmological solution will deviate substantially from the asymptotic solution \eqref{6a} once $\chi$ is of the order of $m$, or $|t|$ of the order $\mu^{-1}$. The leading contribution to $H$ for large negative $t$ obtains by including the next order in an expansion of $V$ for small $x$. We find 
\be\label{6A}
x=\frac{2(6-A)}{(2-A)^2}
\left\{\frac{1}{(1-\mu t)^2}+\frac{c}{(1-\mu t)^4}\right\},
\ee
with 
\be\label{6B}
c=\frac{2(6-A)^2(4-3A)}{A(2-A)^2(10-3A)}.
\ee
One infers an asymptotically vanishing negative Hubble parameter \eqref{2AA} with $h>0$,
\be\label{6C}
h=\frac{4(6-A)^2}{A(2-A)^2(10-3A)},
\ee
such that the universe is slowly shrinking. The scale factor approaches in the infinite past a constant value $\bar a$,
\be\label{6D}
a=\bar a \exp 
\left\{-\frac{h}{2(1-\mu t)^2}\right\}\approx \bar a-
\frac{\bar a h}{2(1-\mu t)^2}.
\ee

\subsection{Regular geometry}
For Minkowski space in the infinite past there is no doubt that geometry is regular. All geodesics are complete towards the past. Of course, the time distance to the ``infinite past'' should be measured with a concept of time that is coordinate invariant, rather than a particular time coordinate. For a pure geometrical concept of time we could take the proper time $\tau$ on time-like geodesics. Indeed, the proper time elapsed since the infinite past is infinite, $\tau(t\to-\infty)\to-\infty$. For the Robertson-Walker metric the time coordinate $t$ actually coincides with the proper time for observers that are at rest in comoving coordinates. With this interpretation it is a reasonable coordinate-invariant time unit. (We will discuss below concepts of ``physical time'' that differ from ``geometrical time''.) Massless particles move on light-like geodesics in the geometry \eqref{6D}. 

\subsection{Asymptotic gravity}
Despite the vanishing of $\chi$ for $t\to-\infty$ the long distance gravitational attraction between massive particles remains weak in this limit, provided particle masses are sufficiently small as compared to $\chi,m_p=h_p\chi$, $h_p\ll 1$. The dimensionless strength 
of the gravitational interaction between massive particles is given by $m^2_p/\chi^2=h^2_p$, and therefore independent of $\chi$.
 
For a discussion of graviton scattering in the limit $\chi\to 0$ we first extend our model by adding to eq. \eqref{1a} a term 
\be\label{11A}
\Delta \Gamma=C\int_x\sqrt{g}R^2.
\ee
The constant $C$ is dimensionless such that $\Delta \Gamma$ is scale invariant, in accordance with a fixed point at $\chi=0$. For $\chi=0$ only this term survives for the pure gravitational self-interaction. For the cosmological solution \eqref{2AA}, \eqref{6A} the asymptotic ratio 
\be\label{6E}
\frac{R}{\chi^2}\approx\frac{6\dot H}{\chi^2}\sim \frac{\mu^2}{m^2}(1-\mu t)^{-\frac{4(1-A)}{2-A}}
\ee
vanishes for $t\to-\infty$, such that higher order invariants as $R^2$ or $R_{\mu\nu }R^{\mu\nu}$ are negligible as compared to $\chi^2 R$. (Note $H^2/\dot H\sim (1-\mu t)^{-2}\to 0$.) The influence of the term \eqref{11A} on the cosmological solution can therefore be neglected, justifying its omission for most parts of this paper. 

For graviton-graviton scattering with transferred squared momentum $q^2$ we have to distinguish two regimes. For $|C q^2|\ll\chi^2$ the contribution of eq. \eqref{11A} is subleading and the cross section is of the form
\be\label{12A}
\sigma\sim \frac{q^2}{\chi^4}\quad \text{ for }\quad|C q^2|\ll\chi^2.
\ee
(Here $q^2$ denotes generically a combination of transferred momenta in different channels. For our considerations the details do not matter - only powers of $\chi$ are relevant and dimensional analysis provides the associated powers of $q^2$.) On the other hand, for $|Cq^2|\gg\chi^2$ the contribution from the term $\sim\chi^2R$ in eq. \eqref{1a} becomes subleading, and $\chi$ no longer appears in the leading contribution to the cross section
\be\label{12B}
\sigma\sim \frac{1}{C^2 q^2} \quad\text{ for } \quad |Cq^2|\gg \chi^2.
\ee
In the transition region $|Cq^2|\approx \chi^2$ the expressions \eqref{12A} and \eqref{12B} match. For any finite $q^2$ and $\chi\to 0$ graviton-graviton scattering will be dominated by eq. \eqref{12B} and become independent of $\chi$, in contrast to the diverging expression \eqref{12A}. 

These considerations also apply to the gravitational attraction between massive particles in the limit $\chi\to 0$ if we consider non-zero $q^2$. The appropriate model for the infinite past describes massless particles coupled to scale invariant higher order gravity. The precise form of the gravitational interaction at the fixed point for $\chi=0$ is not known. We use the simple form \eqref{11A} here only in order to provide for a simple, consistent, and qualitatively correct picture. 

We conclude that our model admits a regular description for the infinite past. It also remains regular in the infinite future (see below) and therefore describes an eternal universe. The infinite past is characterized by flat space with $\chi=0$. For this state all particle masses vanish. One can, of course, construct dimensionless quantities that diverge or vanish in the limit $\chi\to 0$, as the diverging ratio $V/\chi^4$ or the vanishing dimensionless time interval $d\tilde \tau=\chi d\tau$. The latter measures time in units of the diverging inverse particle masses. It seems therefore not surprising that the distance in $\tilde \tau$ from the infinite past to some finite time $t_0$ remains finite. For particles that become massless the inverse particle mass is simply not a very appropriate time unit. One rather may count oscillations of the wave function for finite momenta, similar to photons, cf. sect. \ref{Physical time}.

We emphasize that our model is simple, stable (no ghosts or tachyons) and has attractive gravity. (For a recent debate on geodesic completeness in models of antigravity \cite{Lin} see refs. \cite{BS,KA}.) The strength of the long distance gravitational attraction between massive particles is time-independent. We have found no sign of inconsistency of this model. We will see in sects. \ref{Singularities in the Einstein frame}, \ref{Eternal Universe and inflation} that despite the unusual features the primordial cosmology describes the physics of inflation and generates the primordial density fluctuations, with spectral index $n=1-A$, and large tensor ratio $r=8(1-n)$. 

\section{Focus property of primordial \\cosmology}
\label{Focus property of primordial cosmology}

\subsection{General homogeneous isotropic solutions}
\begin{figure}[h!tb]
\centering
\includegraphics[scale=0.6]{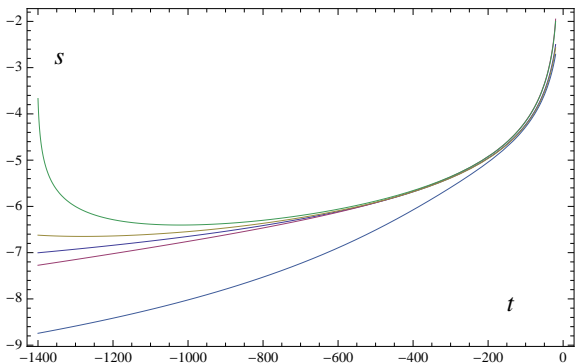}
\caption{Focus properties of primordial cosmology. We show the scalar field $s=\ln(\chi/m)$ as a function of cosmic time $t$ for various initial conditions. The second curve from below is the attractor solution \eqref{6A}. Solutions with arbitrary initial conditions approach this solution as $t$ increases, provided the time scale is suitably adjusted by a linear shift in $t$. Parameters are $A=0.04,\mu=1$. 
}
\label{ETU1}
\end{figure}

A one-parameter family of solutions similar to eqs. \eqref{2AA}, \eqref{6A} obtains by constant shifts of $t$. These solutions are stable. The most general solution of eqs. \eqref{3a}, \eqref{5a} has two integration constants, however. For large negative $t_{\rm in}$ one may specify initial values $\chi_{\rm in}$ and $\dot \chi_{\rm in}$ (or, equivalently, $H_{\rm in}$). For small enough initial values one finds that the general solution is attracted for increasing $t$ towards the family of asymptotic solutions \eqref{6A}. This can be seen in FIG. \ref{ETU1} which shows the time evolution of $s$ for different initial conditions. Starting at $t\ri$ with initial values given by the solution \eqref{6A} the numerical result is indistinguishable from the analytic curve \eqref{6A} in the range shown. For other initial conditions the solutions are attracted towards this universal scaling solution (up to a linear shift in $t$). 

The attractor property of the solution \eqref{6A}, which is characteristic for stable solutions, has an important consequence. At some time $t_0>\ti$ the range of values for $s(t_0)$ and $\dot s(t_0)$, that can be reached for arbitrary initial values at $\ti$, is restricted. The solution of the field equation maps the range $-\infty<s(\ti)<\infty,-\infty<\dot s(\ti)<\infty$ to a finite ``allowed region'' for the values $\big(s(t_0), \dot s(t_0)\big)$. We may associate this with a ``focus property'' of a system of differential equations. 

Inversely, only values of $\big(s(t_0),\dot s(t_0)\big)$ within the allowed region can be continued backwards to $\ti$. For all other values $\big(s(t_0),\dot s(t_0)\big)$ outside the allowed region the solution must diverge somewhere in the interval $\ti<t<t_0$. The focus property of a regular attractor solution implies that there exists a region in the space of solutions that must become singular in the past \cite{CWVG}. On the other hand, arbitrary values of $\big(s(t_0),\
\dot s(t_0)\big)$ lead to regular solutions for the future, $t>t_0$. This asymmetry between the past and the future is due to the ``arrow of time'' generated by the spontaneous breaking of time reversal symmetry by cosmological solutions \cite{CWVG}. One recognizes this feature for the ``bounce solution'' in FIG. \ref{ETU1} (upper curve), for which $s$ first decreases and subsequently increases. This solution diverges for $t$ only somewhat smaller than the range shown in the figure. 

Consider now the limit $\ti\to-\infty$, with initial values specified in the infinite past. The allowed region for finite $t_0$ is constituted by a ``manifold of fixed points''. In our case this manifold is one dimensional. It consists of the family of the universal attractor solutions approximated by eq. \eqref{6A}, with a shift in $t$ as the free parameter. All initial conditions at $t_0$ neighboring one of the universal scaling solutions will lead to solutions diverging in the past. (This holds for an arbitrarily small but nonzero distance from the family of attractor solutions.) 

The focus property of our isotropic and homogeneous solutions is likely to extend to anisotropic and inhomogeneous solutions. Typically, anisotropies and inhomogeneities are ``washed out'' by an inflationary cosmology. We have not yet done the stability analysis of our solution in a more general space of inhomogeneous solutions surrounding it. We only remark here that it is well possible that the focus property is sufficiently strong such that at $t_0$ only the homogeneous and isotropic solution is allowed in the limit $\ti\to-\infty$. In this case we expect that all neighboring anisotropic and inhomogeneous solutions become singular in the past. We conclude that in case of focus properties of an attractor solution the presence of neighboring solutions that diverge in the past is not a sign for a ``beginning'' of the universe. It rather reflects the ``loss of memory'' characteristic for attractor solutions. 

\subsection{Particle trajectories}
The focus property of our cosmological solution is also reflected in the motion of massive particles. The trajectories $x^\mu(\tau)$ of massive particles obey \cite{AWW,BPAW}
\be\label{11n}
\frac{du^\mu}{d\tau}+\Gamma^\mu_{\rho\sigma}u^\rho u^\sigma+\partial^\mu\ln m+u^\mu u^\rho\partial_\rho\ln m=0,
\ee
with $u^\mu=\frac{dx^\mu}{d\tau}$. The usual geodesic equation is modified by the two last terms which reflect the $\chi$-dependence of the particle masses
\be\label{12n}
\partial_0\ln m=\frac{\dot\chi}{\chi}~,~\partial_k\ln m=0.
\ee
The nonzero elements of the connection for a Robertson Walker metric are $\Gamma^0_{ij}=Hg_{ij}=Ha\delta_{ij},\Gamma^j_{0i}=H\delta^j_i$. The direction of the velocity $u^k$ does not change and we denote by $u$ the length of $u^k$. With $u^0=\gamma$ and $x^0=t$ the trajectories of massive particles are given by
\be\label{GX1}
\frac{\partial u}{\partial \tau}=-(2H+\dot s)\gamma u~,~
\frac{\partial\gamma}{\partial\tau}=-(H+\dot s)(\gamma^2-1),
\ee
where we use the definition of proper time, $\gamma^2=1+a^2u^2$. Proper time is related to the time coordinate in the Robertson-Walker metric by
\be\label{15A}
\frac{dt}{d\tau}=\gamma.
\ee
The varying mass adds substantial complication to the use of proper time for physical time measurements. Massive particles behave as photons in the infinite past where their mass $\sim\chi$ vanishes. We will discuss this issue by detailed solutions of eq.\eqref{GX1} since it will be crucial for the interpretation of incomplete geodesics in the Einstein frame. 

The term $\dot s$ dominates over $H$ for $t\to-\infty$. For the asymptotic behavior we neglect $H$ and approximate $s$ by eq. \eqref{6a},
\be\label{GX2}
\dot s=\frac{2\mu}{(2-A)(1-\mu t)}~,~\frac{\partial u}{\partial t}=-\dot s u.
\ee
The solution 
\be\label{GX3}
u(t)=u(\ti)
\left(\frac{1-\mu t}{1-\mu\ti}\right)^{\frac{2}{2-A}}
\ee
shows a strong focus property. The increase of the mass damps velocities by a factor $(t/t_{in})^{2/(2-A)}$ (for $|t|\gg \mu^{-1})$. Since the evolution of $a$ is a subleading effect this extends to $\gamma$,
\be\label{GX4}
\gamma^2(t)-1=\big(\gamma^2(\ti)-1\big)
\left(\frac{t}{\ti}\right)^{\frac{4}{2-A}}.
\ee
We observe that particles at rest (in comoving coordinates) are singled out as a fixed point, $u=0,\gamma=1$. If we start at $\ti$ with some finite maximal value $\gamma_{\max}$ the particle trajectories are attracted towards this fixed point. For $\ti\to-\infty,~\gamma(\ti)\leq\gamma_{\max}$ all particles have come to rest at finite $t_0$, $\gamma(t_0)=1$. 

On the other hand, the physical momentum $p\sim\chi {au}$ remains constant in leading order, the decrease of $u$ being canceled by the increase of the mass $\sim\chi$, cf. eq.  \eqref{6a}. A change of $p$ arises only in next to leading order. As a consequence of translation symmetry one has a conserved quantity, $ap=a^2\chi u=$const., such that $p\sim a^{-1}$. 

\section{Physical time}
\label{Physical time}

\subsection{Proper time on particle trajectories}
We may evaluate the proper time on a trajectory of a massive particle rather than on a time-like geodesics. For particles at rest these two concepts coincide. For moving particles, however, one finds on a particle trajectory \eqref{GX4}  $\big (\gamma\ri=\gamma(\ti)\big)$
\be\label{GX5}
\frac{dt}{d\tau}=\gamma=
\sqrt{(\gamma^2\ri-1)
\left(\frac{t}{\ti}\right)^{\frac{4}{2-A}}+1}.
\ee
For $\gamma\ri<\gamma_{\max}$ and $\ti\to-\infty$ this yields $\tau=t+c$, such that proper time diverges in the infinite past. 

In contrast, for particles with non-vanishing momentum $p(t)$ one finds a finite proper time distance to the infinite past. This reflects that such particles behave as photons in the infinite past such that proper time is no longer a valid concept for physical time, see below. For particles with constant non-zero momentum $p$ one has $\gamma^2\ri-1\sim p^2/\chi^2\ri\sim\ti^{4/(2-A)}$, such that 
\be\label{GX6}
\frac{dt}{d\tau}=\sqrt{1+|\nu t|\f},
\ee
with constant $\nu\sim p^2$. One finds two regimes. As long as particles are relativistic one has $|\nu t|\gg 1$ and 
\be\label{GX7}
\tau-\tau\ri=\frac{2-A}{A\nu}
\left((-\nu t)\f-(-\nu \ti)\f\right).
\ee
This difference remains finite for $\ti\to-\infty$. If particles become non-relativistic at $t_{\rm nr}$ due to their increasing mass they enter the regime $|\nu t|\ll 1$ for which $\tau-\tau\nr\approx t-t\nr$. Trajectories of particles with non-zero $p$ always belong to the relativistic regime in the past. Since $t_{nr}$ is finite there exists always a period where $|t|\gg |t\nr|$. In consequence, the proper time elapsed on those trajectories between the infinite past $\ti\to-\infty$ and finite $t_0$ remains finite.

If one would measure time intervals in units of the inverse particle mass, $d\tilde{\tau}=\chi d\tau$, one would find a finite distance to the infinite past for {\em all} trajectories, including the ones for particles at rest. This reflects that the inverse particle mass $\sim\chi^{-1}$ is no longer a suitable time unit for $\chi\to 0$. In order to avoid the complications of a photon-like behaviour the use of proper time for a definition of ``physical time'' should be restricted to trajectories with finite $\gamma$ and time units given by $\mu^{-1}$. In this case the distance to the infinite past turns out to be indeed infinite.

\subsection{Oscillation time}
These findings do not indicate the necessity of a ``beginning'' of the universe. They rather remind us that proper time is not an appropriate measure of time for photons. For photons with given comoving wave vector $k$ (which is proportional to momentum in our case in the limit $t\to-\infty$) a useful coordinate invariant measure of time is given by the number of oscillations of the field amplitude. The counting of oscillations is best done by use of conformal time, $ds^2=a^2(\eta)(-d\eta^2+d\vec x^2),~d\eta=dt/a$. The wave equation for a particle with mass $m$ reads in conformal time 
\be\label{PT1}
(\partial^2_\eta+2Ha\partial_\eta+k^2+a^2m^2)\varphi_k=0,
\ee
where we consider for definiteness a complex scalar field mode $\varphi_k$ with $k$ the comoving wave vector. Using $H=\partial_t\ln a =a^{-2}\partial_\eta a$, one can factor out the Hubble damping of the amplitude
\be\label{PT2}
\varphi_k=\frac{\tilde\varphi_k}{a}~,~\left\{\partial^2_\eta +k^2+a^2\left (m^2-\frac R6\right)\right\}\tilde\varphi_k=0.
\ee
Towards the infinite past $R$ vanishes, such that for $m=0$ the oscillations are simply given by
\be\label{PT3}
\tilde\varphi_k=\exp(-ik\eta)\tilde\varphi_{k,0}.
\ee
The number of oscillations $n_k$ is proportional to conformal time 
\be\label{PT4}
n_k=\frac{k\eta}{\pi}.
\ee

Let us define dimensionless ``oscillation time'' or ``period time'' $\tilde t_p$ by the number of zeros of a given component of the wave function. (In our case we may take the real part of $\varphi_k$, for photons it could be a given component of the electric or magnetic field.) The number of zeros of the wave function does not depend on the choice of coordinates such that $\tilde t_p$ is a coordinate - invariant quantity
\be\label{25A}
\tilde t_p=n_k.
\ee
If we associate different clocks to different $k$-modes they tick with different frequences. The different dimensionless oscillation times $\tilde t_p$ can be gauged to a common oscillation time $t_p$ with dimension of length by a suitable normalization, see below. For massless particles the oscillation time is proportional to conformal time for our choice of coordinates, $t_p\sim\eta$. The distance in oscillation time diverges towards the infinite past $t\sim\eta\to-\infty$. Wave functions have undergone an infinite number of oscillations since the infinite past. 

Oscillation time can be defined for massive particles as well. In the limit $k^2\ll a^2m$ one has 
\be\label{27A}
\left(\partial^2_\eta+a^2\left(m^2-\frac{R}{6}\right)\right)\tilde\varphi=0,
\ee
or
\be\label{25B}
\left(\partial^2_t+H\partial_t+m^2-\frac R6\right)\tilde\varphi=0.
\ee
Correcting for the different quantitative role of the Hubble damping yields 
\be\label{25C}
(\partial^2_t+m^2-\frac94H^2-\frac32\dot H)\bar\varphi=0~,~
\bar\varphi=a^{1/2}\tilde\varphi=a^{\frac32}\varphi.
\ee
For constant $m^2$, and if $H^2$ and $\dot H$ can be neglected as compared to $m^2$, one obtains the oscillation
\be\label{25D}
\bar\varphi=\exp (-imt)\bar\varphi_0,
\ee
with the number of zeros
\be\label{25E}
n=\frac{mt}{\pi}=\tilde t_p.
\ee
In this limit the oscillation time is proportional to the proper time for particles at rest (in comoving coordinates).

If mass is constant and the modifications from geometry $(\sim H^2,\dot H)$ are subleading, the oscillation time interpolates between proper time for massive particles at rest and conformal time for photons. More generally, oscillation time is defined for arbitrary $k$, arbitrary varying scale factor $a(t)$ or varying mass $m(t)$. Furthermore, it can be extended beyond the isotropic and homogeneous cosmological solutions to geometries with arbitrary metric and masses with arbitrary dependence on space and time. The only requirement for a suitable ``clock'' for which this time is defined and measured is the occurrence of ``regular zeros'' in the components of the wave function. This holds for all periodic processes, but strict periodicity is not required. Clocks can be defined for quantum systems with $\varphi$ the wave function, or classical fields as electromagnetic fields. This universality of oscillation time makes it a robust candidate for physical time also for situations where massive particles behave as photons and proper time can no longer be used. 

Oscillation time can be related directly to physical time measurements. For example, atomic clocks count the number of oscillations of the non-relativistic wave function. (This replaces $m$ effectively by an energy difference for different atomic levels, which is proportional to the electron mass $m$ and will vary with $m$ in our type of cosmology.) Oscillation time keeps a meaning even if observers performing measurements no longer exist - oscillating fluctuations still play a role for inflationary cosmology. A counting of discrete zeros does not depend on the system of coordinates. Furthermore, field redefinitions by multiplication with a non-zero function, as the Weyl scaling between the freeze and Einstein frames, do not change the number of zeros. We conclude that $\tilde t_p$ is coordinate {\em and} frame invariant. In the following we will associate the oscillation time with {\em physical time}.

The transition from discrete counting to a continuous dimensionless time variable is straightforward if the discrete time intervals are small as compared to the elapsed time. We still need a normalization that multiplies $\tilde t_p$ with an appropriate length scale. In a translation invariant setting modes with different $k$ do not mix if the amplitude is small enough such that the linear field equation \eqref{PT1} applies. For massless particles we choose the normalization
\be\label{PT5}
t_p=\frac{\pi \tilde t_p}{k},
\ee
where $k$ is an arbitrary wave number different from zero. With this normalization all clocks with arbitrary non-zero $k$ measure the same time. For photons $t_p$ coincides with $\eta$ in the infinite past. In turn, with the choice $\bar a=1$, conformal time $\eta$ equals $t$, confirming that the distance in physical time to the infinite past is indeed infinite. 

For massive particles we may take for time intervals the normalization
\be\label{26A}
dt_p=\frac{\pi d\tilde t_p}{\sqrt{k^2+a^2m^2}},
\ee
which coincides with eq. \eqref{PT5} for $k^2\gg a^2m^2$. In the opposite limit $k^2\ll a^2m^2$ it yields an expression that is independent of $m$,
\be\label{27B} 
dt_p=\frac{dt}{a}~,~t_p=\eta.
\ee
We have formulated the normalization of $t_p$ here in a given coordinate system and frame. It has to be transformed appropriately to other coordinates or frames. One may try to find a direct coordinate-invariant expression for the normalization. This is not of major concern in the present context, since the most relevant quantity is the coordinate- and frame-invariant dimensionless time $\tilde t_p$. For practical purposes one can often associate physical time $t_p$ with conformal time $\eta$, but one should recall that the basic definition of physical time is the counting procedure for $\tilde t_p$. 

\subsection{Limits for proper time}
While the concept of physical time is robust enough to cover both photons and massive particles, this does not hold for proper time. Particles with non-zero $p$ always behave as photons for $t\to-\infty$ since the mass vanishes in this limit. Proper time is therefore not suitable for a time measurement. In other words, the use of proper time for a time measurement should be restricted to trajectories for which $\gamma$ remains finite. These are the ones with finite velocities, or finite $\gamma\ri$ for $t\ri\to-\infty$. For those $\tau(t_0)-\tau\ri$ is indeed infinite for $\ti\to-\infty$ and finite $t_0$. 

We summarize that a finite distance to the infinite past occurs only for quantities that do not constitute suitable time measurements, as proper time $\tau$ for trajectories of particles that become photon-like for $t\to-\infty$, or proper time $\tilde \tau$ in units of the inverse particle mass, $d\tilde\tau=\chi dt$. In both cases the clock stops in the infinite past, while better ``physical clocks'' continue to show regular and finite time intervals. It will be precisely those ``stopping clocks'' that are reflected by the incomplete geodesics in the Einstein frame. We will see in the next section that proper time is not a frame-invariant concept.

\section{Singularities in the Einstein frame}
\label{Singularities in the Einstein frame}

\subsection{Big bang frame and freeze frame}
Our family of models \eqref{1a}, \eqref{2a} can be described in a different ``big bang picture'' by performing a Weyl scaling to the Einstein frame,
\be\label{6F}
g_{\mu\nu}=\frac{M^2}{\chi^2}g'_{\mu\nu}~,~\sigma=\sqrt{B}M\ln\frac{\chi}{M}.
\ee
In the new ``field coordinates'' $g'_{\mu\nu}$ and $\sigma$ the effective action describes a standard setting with constant Planck mass $M$, 
\be\label{6G}
\Gamma=\int_x\sqrt{g'}\left\{-\frac12M^2R'+\frac12\partial^\mu\sigma \partial_\mu\sigma +V'(\sigma)\right\}.
\ee
Particle masses $m=h_pM$ are constant as well. The potential, $V'=(M^4/\chi^4)V$, decays exponentially for large $\sigma$, 
\ba\label{6H}
V'&=&\frac{\mu^2M^4}{m^2
\left[\exp \left(\frac{\alpha\sigma}{M}\right)+\exp 
\left(\frac{\tilde\alpha\sigma}{M}\right)\right] },\nn\\
\alpha&=&\frac{2}{\sqrt{B}}~,~\tilde\alpha=\frac{A}{\sqrt{B}}.
\ea

For our first model with $B=A$ we concentrate on small $A<0.04$, such that $\alpha>10,~\tilde \alpha<0.2$. For primordial cosmology one has $\sigma\to-\infty$, such that the term involving $\tilde \alpha$ dominates in $V'$. This potential describes power-law inflation \cite{AW}, with spectral index $n\approx 1-\tilde \alpha^2$ and large tensor amplitude $r\approx 8\tilde\alpha^2=8(1-n)$. In the Einstein frame the Hubble parameter diverges in the ``extreme past'' $t\to 0$,
\ba\label{6I}
H_E&=&\frac{2}{\tilde\alpha^2 t},\nn\\
\sigma&=&\frac{2M}{\tilde \alpha}
\left\{ \ln \left(\frac{M\mu t}{m}\right)-\ln
\sqrt{\frac{12-2\tilde \alpha^2}{\tilde \alpha^4}}\right\},
\ea
such that the curvature scalar $R'$ and other invariants diverge. The proper time between the extreme past and some finite time $t_0>0$ is finite, all timelike geodesics are incomplete towards the past. The geometry becomes singular for $t\to 0$, and this has led to the judgment that the universe of such a model cannot be eternal and must have had some beginning. 

In view of the regularity of our model in the ``freeze frame'' \eqref{1a}, \eqref{2a} this singularity finds a different interpretation. It is a pure property of the choice of fields $g'_{\mu\nu},\sigma$ which becomes singular for $\chi\to 0$, cf. eq. \eqref{6F}, rather than being connected to a physical singularity. This demonstrates our central point (i) on the possible occurrence of field singularities. Observations are independent of the choice of frame. Indeed, the scale factor, Hubble parameter and proper time in the Einstein and freeze frames are related by
\be\label{6J}
a_E=\frac{\chi}{M} a_{f}~,~H_E=\frac{M}{\chi}\left(H_{f}+\frac{\dot\chi}{\chi}\right)~,~d\tau_E=\frac\chi Md\tau_{f},
\ee
with time derivative $\dot\chi$ taken in the freeze frame. With these transformations observables can be mapped from one frame to the other. 

The relations \eqref{6J} are most easily obtained if we use conformal time, $g_{\mu\nu}=a^2(\eta)\eta_{\mu\nu}$. Then the coordinates $(\eta,\vec x)$ and the scalar field $\chi(\eta)$ are invariant under the conformal transformation \eqref{6F}, while the scale factor transforms according to the first equation \eqref{6J}. The relation for the Hubble parameters follows by computing $\partial \ln a/\partial t=a^{-1}\partial\ln a/\partial\eta$ in both frames. Finally, with invariant $(\eta,\vec x)$ the relation for proper time follows from the defining relations
\be\label{XAX}
d\tau^2_E=a^2_E(d\eta^2-d\vec x^2)~,~d\tau^2_f=a^2_f(d\eta^2-d\vec x^2).
\ee
We emphasize that proper time is not invariant under Weyl scaling. Similarly, geodesics in one frame are not mapped to geodesics in another frame, except for massless particles and massive particles at (comoving) rest.  The last equation \eqref{6J}, together with the coordinates $x^k$ being kept fixed, yields the frame transformation for the velocity $u$ and momentum $p=mau$,
\be\label{32A}
u_E=\frac M\chi u_f~,~m_E=\frac M\chi m_f~,~p_E=\frac M\chi p_f~,~\tilde p_E=\tilde p_f.
\ee
The momentum divided by mass, $\tilde p=p/m=au$, is a frame invariant quantity.  

The transformation rules \eqref{6J}, together with the invariance of $\chi(\eta)$, allow us to transform any cosmological solution in the freeze frame to the Einstein frame and vice versa. Equivalently, we may solve the field equations derived from the effective action \eqref{6G} without any reference to the freeze frame. One can then verify that these solutions are mapped to solutions of the field equations of the freeze frame by the transformation \eqref{6J}. Eq. \eqref{32A} specifies how velocity, mass and momentum on a particle trajectory are mapped from one frame to the other. We have employed here the particular conformal transformation \eqref{6F}. For the general expression one simply replaces $M/\chi$ by the appropriate conformal factor. 

We emphasize again that the dimensionless oscillation time $\tilde t_p$ is based on a discrete counting of zeros which is the same in all frames. This physical time is frame invariant. For the frames that we consider in this paper also conformal time $\eta$ is frame invariant. This comes in pair with the (approximate) association of normalized physical time $t_p$ with $\eta$ in eqs. \eqref{PT5}, \eqref{27B}. 

The extreme past $t\to 0$ in the Einstein frame corresponds to the infinite past $t\to -\infty$ in the freeze frame. The singularity in $H_E$ is a simple consequence of the transformation \eqref{6J}. The finite proper time towards the extreme past in the Einstein frame reflects the finite value of $\tilde \tau$ in the freeze frame, that we have discussed before. Timelike geodesics in the Einstein frame are mapped to trajectories of massive particles in the freeze frame. Physical observables typically involve dimensionless ratios and do not depend on the choice of frame. Since we have already established a frame where observables remain regular from the infinite past to the infinite future the universe of this model is eternal. The Einstein frame is simply poorly adapted to the asymptotic situation where the Planck mass and particle masses vanish. The singularity in the Einstein frame reflects the inappropriate choice of time or length measured in units of inverse particle masses. In the remaining part of this section we discuss this issue in more detail.

\subsection{Limits for the use of proper time}
The explicit map between the freeze and Einstein frames may be used to develop a new view on the meaning of the singularities in the Einstein frame. They are not connected to singularities for physical observables, but rather to insufficiencies of certain concepts for a definition of useful physical observables. A first example is the definition of a physical observable for time. For massive particles, proper time in units of the inverse particle mass, evaluated on the particle trajectory, is a frame invariant concept and therefore a priori a reasonable candidate for physical time. Indeed, one has $d\tilde \tau_f=Md\tau_E$ and particle trajectories in one frame are mapped to particle trajectories in any other frame. The use of proper time for a definition of physical time intervals is restricted,however, to particles with a non-zero ratio mass/momentum, while it becomes inappropriate for photons or a photon-like behavior for which this ratio vanishes. Following in the Einstein frame the trajectories of 
massive particles with non-zero momentum 
towards the past one finds that the ratio mass/momentum reaches zero at the singularity. In the vicinity of the singularity proper time is no longer a valid concept for physical time. 

The inappropriateness of proper time for the vicinity of the big bang is visible in the Einstein frame without any reference to the freeze frame. While particle masses are constant in the Einstein frame, the momentum diverges, such that the relevant ratio $m^2/p^2$ reaches zero for $t\to 0$. All particles with non-zero $p(t_0)$ at some non-zero time $t_0$ become photon like for $t\to 0$. For the oscillations of the associated field \eqref{PT1} one always enters the regime $k^2\gg a^2 m^2$ for $t\to 0$.

In contrast, the definition of physical time by the number of oscillations of the wave function can be extended to the singularity. For photons in an isotropic and homogeneous geometry with $a^2R\to 0$ this definition of physical time coincides with conformal time $\eta$, cf. eq. \eqref{PT5} . For the solution \eqref{6I} one has $(\tilde \alpha^2=A<2)$
\ba\label{CT1}
a_E(t)=c_at^{\frac{2}{\tilde\alpha^2}}~,~R_E=\frac{36}{\tilde \alpha^2 t^2}~,~
~\eta(t)=-c_\eta t^{1-\frac{2}{\tilde\alpha^2}},
\ea
such that $a^2R$ indeed vanishes for $t\to 0$. Conformal time diverges for $t\to 0,~\eta(t\to 0)\to-\infty$. In these oscillation time units the extreme past $t\to0$ is at infinite time-distance, as expected for the infinite past in the freeze frame. This extends to massive particles with non-zero $p(t_0)$ for which $k^2\gg a^2m^2$ becomes valid close to the big bang. Physical time measured by the number of oscillations in their wave function becomes again proportional to conformal time in the extreme past. Again the physical time distance diverges for $t\to 0$, in contrast to proper time. 

We conclude that for the power-law inflation in the Einstein frame \eqref{6I} proper time is not a valid concept for a definition of physical time near the big bang, while for physical time as defined by the number of oscillations in the wave function no singularity occurs for any finite time distance in the past. We note that the asymptotic behavior of the scalar field according to eq. \eqref{6I},
\be\label{CT2}
\chi(\eta\to-\infty)=c_\chi(-\eta)^{-\frac{2}{2-\tilde\alpha^2}},
\ee
coincides with the one computed in the freeze frame, eq. \eqref{6a}, as it should. 

\subsection{Ambiguity of geometry}
Let us next turn to the geometric singularities present in the Einstein frame for the extreme past $\eta\to-\infty$. They occur at infinite distance in physical time, while for any finite $\eta$ the geometry remains regular. We should emphasize, however, that geometry is not a universal concept. It depends on the choice of the metric - different choices can lead to very different geometries. A priori it is not clear if one can single out a ``physical geometry'' and what should be the precise criterion for this purpose. This issue concerns the observation that one can form many combinations of fields that transform as symmetric second rank tensors and are therefore candidates for a metric, permitting the construction of an associated geometry. If the non-vanishing bosonic fields besides the graviton correspond all to massive particles one may formulate a concept of ``universal geometry'' at least for distances larger than $m^{-1}_{\min}$, with $m_{\min}$ the smallest boson mass \
\cite{CWUG}. In general, the vicinity of the big bang typically involves distances smaller than $m^{-1}_{\min}$ and the universality of geometry breaks down. Our models contain an almost massless scalar field such that a universal geometry cannot even be expected for distances that are characteristic for late cosmology.

The singularities occur if we construct the curvature scalar and similar invariants from the Einstein-frame metric $g'_{\mu\nu}$. One may propose a different geometry, with metric
\be\label{CT3}
g_{\mu\nu}=\exp \left(-\frac{2\sigma}{\tilde\alpha M}\right)g'_{\mu\nu}.
\ee
Employing $g'_{\mu\nu}=a^2_E(\eta)\eta_{\mu\nu}$, and
\be\label{35A}
\exp \left(\frac{\sigma(\eta)}{\tilde\alpha M}\right)=\frac{\chi(\eta)}{M}=
\frac{a_E(\eta)}{\bar a}, 
\ee
according to the asymptotic solution \eqref{CT2}, \eqref{CT1}, one finds that the geometry constructed from $g_{\mu\nu}$ is flat space, $g_{\mu\nu}=\bar a^2\eta_{\mu\nu}$. (This corresponds to the lowest order solution \eqref{6a} in the freeze frame, $g_{\mu\nu}$ and $g'_{\mu\nu}$ being related by eq. \eqref{6F}.) The same ambiguity for the choice of metric exists, of course, if we start from the freeze frame. Computing the curvature scalar $R'$ for the metric $g'_{\mu\nu}=(\chi^2/M^2)g_{\mu\nu}$ one finds a diverging $R'$ for the infinite past, $t=\eta\to-\infty$. 

The choice of metric and geometry is closely linked to the choice of a unit in which length is measured. The metric $g'_{\mu\nu}$ corresponds to a length unit given by the inverse particle mass, while $g_{\mu\nu}$ uses a fixed scale as $\mu^{-1}$ as a length unit. In the freeze frame the origin of the singularity in $R'$ is easily understandable. If the inverse particle mass $\chi^{-1}$ diverges in the infinite past, every length measured in this unit shrinks to zero, such that the geometry becomes singular. Nothing forces one, however, to employ this singular geometry for the description of observations. For the infinite past the regular metric $g_{\mu\nu}$ seems better suited for a formulation of useful geometrical concepts. 

\subsection{Physical observables}
Physical observables do not depend on the choice of frame. If expressed in terms of frame-invariant quantities we can evaluate them in all frames by the same prescription. (Otherwise observables have to be transformed under frame changes, cf. eq. \eqref{6J}. This is similar to transformations under diffeomorphisms which can also be formulated as field transformations.) A few frame-invariant quantities are easily established if the field transformation does not involve derivatives, as in eq. \eqref{6F}. In this case terms in the Lagrangian with different numbers of derivatives are not mixed by the frame transformation. In our case this concerns the potential term ${\cal L}_0=\sqrt{g}V=\sqrt{g'}V'$ and the combined kinetic term 
\ba\label{CT5}
&&{\cal L}_{2}=\sqrt{g}
\left\{-\frac12\chi^2 R+\frac12(B-6)\partial^\mu\chi\partial_\mu\chi\right\}\nn\\
&&~ =\sqrt{g'}
\left\{-\frac12 M^2 R+\frac12\partial^\mu\sigma\partial_\mu\sigma\right\}.
\ea
In the freeze frame it is obvious that both ${\cal L}_0$ and ${\cal L}_2$ vanish for the infinite past, since $\chi\to 0,~V\to 0,~R\to 0,~\sqrt{g}\to\bar a^4$. The same holds in the big bang frame. Thus no singularity is present for these frame invariant quantities.

Of course, many other physical observables beyond the explicit frame invariant quantities ${\cal L}_{0,2}$ exists. One example are clocks associated to $\tilde t_p$. We are not aware of any physical observable that becomes singular for our cosmological solution. 

We conclude that the geometric singularities of power law inflation in the Einstein frame can be interpreted as field singularities associated with a particular choice of metric. Physical observables remain regular. In this context we stress that quantities that diverge proportional to powers of $p^2/m^2$ for $m^2/p^2\to 0$ should not be associated with physical observables in a world for which all particles are massless at the big bang. Dimensionless physical observables should not depend on $m$ in this limit. 

Since the singularity in the Einstein frame arises from the singularity of the field transformation for $\chi=0$ one can,in principle, obtain pre-big-bang cosmologies where the singularity is crossed smoothly. It is sufficient that a solution in the freeze frame crosses smoothly the value $\chi=0$. We have, however, not found such a solution for the model \eqref{1a}, \eqref{2a} with $B=A$, at least not within the investigated homogeneous and isotropic setting.  Even if we start with very small positive $\chi$ and very large negative $\dot\chi$ the evolution of the scalar field is strongly damped, the decrease of $\chi$ stops for $\chi_t>0$, and $\chi$ increases subsequently according to the solution \eqref{6A}. 

\section{Crossover model for de Sitter inflation}
\label{Crossover model for de Sitter inflation}

For further illustration of our central point (iv) we discuss a different model \cite{CWSF}, namely $A=0,~B=4/\alpha^2\ll 1$. It describes a crossover between two fixed points for $\chi\rightarrow0$ and $\chi\rightarrow\infty$. For this model the inflationary epoch in the Einstein frame approaches a de Sitter solution in the infinite past. Singularities in geometrical invariants are therefore absent both in the freeze and the Einstein frame, despite the singularity in the conformal transformation. In the existing literature de Sitter space is often called singular for reasons of geodesic incompleteness. Indeed, geodesics of moving massive particles entail a finite proper time when continued to the infinite past. Only for massive particles at rest proper time becomes infinite for $t\to-\infty$. 

We will show that for a non-vanishing momentum $p(t_0)$ at some finite time $t_0$ the momentum $p(t\ri)$ diverges for $t\ri\to-\infty$ both in the freeze- and Einstein frame. This differs from our 
first model for which $p(t\ri)$ remains finite in this limit in the freeze frame. Nevertheless, only dimensionless quantities as $p/m$ matter for observations, and this frame invariant quantity diverges for both models. It may well be possible to find for our second model a further frame where $p(t\ri)$ remains finite, and only $m(t\ri)$ vanishes. 

The basic observation that massive particles with $p(t_0)\neq 0$ become photon-like for $t\ri\to-\infty$, and proper time therefore fails to be a good description of physical time, is the same in both models. The detailed investigation of two different models will reveal, however, that basic general properties can appear in rather different forms in different models. 

\subsection{Infinite past}
In the freeze frame the attractor solution for primordial cosmology can again be extended to the infinite past, where the Hubble parameter vanishes. In contrast to eq. \eqref{6D} the scale factor goes to zero, however,
\be\label{1n}
a(t)=\exp 
\left\{-\left(-\frac{\sqrt{3}\mu t}{2\alpha}\right)^{\frac23}\right\}~,~H(t)=\left(-\frac{2\mu^2}{9\alpha^2t}\right)^{\frac13}.
\ee
The cosmon field $\chi$ increases from its vanishing asymptotic value as
\be\label{4n}
\chi(t)=m
\left(-\frac{2}{\sqrt{3}\alpha^2\mu t}\right)^{\frac13}=
\frac{\sqrt{3}m}{\mu}H(t).
\ee
The validity of this primordial solution breaks down at the end of inflation and we only consider $t< 0$. 

The relative time derivatives vanish for $t\to-\infty$,
\be\label{5n}
\frac{\dot{H}}{H^2}=\frac{\dot{\chi}}{H\chi}=
\left(-\frac{\alpha}{\sqrt{6}\mu t}\right)^{2/3}=-\frac{1}{3Ht},
\ee
in contrast to our first model. The curvature tensor is then approximated by
\be\label{6n}
R_{\mu\nu\rho\sigma}=H^2(g_{\mu\rho}g_{\nu\sigma}-g_{\mu\sigma}g_{\nu\rho}).
\ee
Invariants formed by contracting $n$ factors of $R_{\mu\nu\rho\sigma}$ with $2n$ factors of $g^{\mu\nu}$ vanish $\sim H^{2n}$, e.g. $R=12H^2$. (Subleading contributions are further suppressed by factors of $\dot{H}/H^2$.) For the covariant derivatives one has
\be\label{7n}
D_0R\m=\frac{2\dot H}{H^2}R\m~,~D_k R\m=0,
\ee
such that all invariants involving powers of covariant derivatives vanish as well. We conclude that with respect to all those invariants the limit $t\to -\infty$ remains regular and approaches the properties of flat space.  

\subsection{Proper time for particle trajectories and} 

~{\bf geodesics}

For a determination of the ``time-distance'' to the infinite past we consider first the proper time for massive particles, evaluated on their physical trajectories. With $\dot\chi/\chi=-1/(3t)$ these trajectories obey
\ba\label{13n}
\frac{\partial u}{d\tau}&=&-2H\gamma u+\frac{\gamma u}{3t},\nn\\
\frac{d\gamma}{d\tau}&=&\left(\frac{1}{3t}-H\right)(\gamma^2-1).
\ea
Particles at rest define the``static trajectories'' $u=0,\gamma=1,\tau=t$. The proper time elapsed for particles at rest from the infinite past to some finite $t$ is indeed infinite.

We next establish that for this model all massive particles with finite velocity or momentum approach the static trajectories. The velocity $u$ is damped both by Hubble damping and the increasing mass
\be\label{14n}
\frac{d\ln u}{dt}=-2H+\frac{1}{3t}.
\ee
The solution,
\be\label{15n}
u(t)=u_{\rm in}
\left(\frac{t}{\ti}\right)^{\frac13}
\frac{\exp \left\{\left(\frac{6\mu^2}{\alpha^2}\right)^{\frac13}(-t)^{\frac23}\right\}}
{\exp\left\{\left(\frac{6\mu^2}{\alpha^2}\right)^{\frac13}(-\ti)^{\frac23}\right\}},
\ee
shows the approach of trajectories with arbitrary finite $u_{\rm in}$ towards a static trajectory. For the physical momentum $p=m(\chi)au\sim \chi au$ one finds again $pa=$const. If $p(t)\neq 0$ for finite $t$ the physical momentum diverges in the infinite past, $|p(t\ri \to-\infty)|\to\infty$. Thus any finite $p(t\ri),~t\ri\to-\infty$, is mapped to $p(t)=0$.

The relation between $t$ and the proper time is determined by
\be\label{20n}
\frac{d\tau}{dt}=\frac{1}{\gamma}=
\big(1+a^2(t)u^2(t)\big)^{-\frac12},
\ee
with
\be\label{19n}
a^2(t)u^2(t)=\gamma^2(t)-1
=(\gamma^2_{\rm in}-1)
\frac{a^2_{\rm in}}{a^2(t)}
\left(\frac{t^2}{t^2_{\rm in}}\right)^{\frac13}.
\ee
The proper time difference $\tau-\tau_{\rm in}=\tau(t)-\tau(\ti)$ is a function of $\gamma_{\rm in}$ and $\ti$. We will be interested in the limit $t\ri\to-\infty$. 

The physical momentum of a massive particle, divided by its mass, is given in our coordinate system by $\tilde p=au$, such that $\gamma=\sqrt{1+\tilde p^2}$. Let us call those particles for which $\tilde p$ remains finite in the infinite past ``asymptotic massive particles''. Asymptotic massive particles never behave as photons. Proper time in units of inverse particle mass, evaluated on the trajectory, is a reasonable measure of time. For finite physical momenta in units of particle  mass $\gamma$ is finite.  It is straightforward to show that for an arbitrarily large finite $\gamma_{\rm in}$ the difference $\tau-\tau_{\rm in}$ diverges for $\ti\to-\infty$. The proper time distance to the infinite past is infinite for asymptotic massive particles (criterion (iv) in sect. \ref{Introduction}). Asymptotic massive particles have come to rest at finite time $t$, $u(t)=0$. We may also consider finite momentum $p\sim \chi\tilde p$. This implies that $\tilde p$ increases for $\ti\to-\infty$ at most as $\chi^{-1}$, 
and $\gamma_{\rm in}$ as $\chi^{-2}\sim(-\ti)^{2/3}$. Our conclusion remains unchanged if we allow $\gamma_{\rm in}$ to increase $\sim(-\ti)^{2/3}$, cf. eq. \eqref{19n}. Thus the proper time also diverges for particles with finite momentum in the infinite past. In contrast, the proper time distance to the infinite past remains finite for particles with $u(t)\neq 0$, similar to our first model. 

Proper time for trajectories of massive particles with finite momentum at $\ti$ behaves differently for $\ti\to-\infty$ in our two models. The distance to the infinite past is finite in the first model $(A=B)$ and infinite in the second model $(A=0)$. (In both models particles become photon-like $(\tilde p\to\infty)$ in the infinite past.) For the second model $(A=0)$ the time difference $t\nr-\ti$ for a particle to become non-relativistic is finite, such that $t-t\nr$ diverges for $\ti\to-\infty$. Massive particles behave similar to photons only for a negligible time. Proper time remains a valid concept for most parts of the past trajectory. It diverges for the infinite past. For the first model $(B=A)$ one finds a diverging interval $t\nr-\ti$ for $\ti\to-\infty$, and finite $t-t\nr$. Particles with finite momentum behave photon-like for most of their trajectory, and proper time no longer acts as a useful clock. We recall, however, that finite momentum $p$ is not a frame-invariant concept. The crucial frame invariant quantity is $\tilde p$. For all trajectories with $u(t)\neq 0$ for finite $t$ the dimensionless $\tilde 
p$ diverges for the infinite past in both models. For these trajectories proper time cannot be used towards the infinite past as physical time. The proper time distance to the infinite past can be finite for such trajectories without contradicting the eternity of the universe. 

The proper time for trajectories of massive particles differs from geodesics due to the $\chi$-dependent mass. If we would evaluate the proper time along time-like geodesics the terms $\sim 1/t$ in the two equations \eqref{13n} would be absent. These terms are subleading for $t\to-\infty$, however, suppressed by a factor $1/(Ht)$ as compared to the leading term. The qualitative conclusion for geodesics is the same as for particle trajectories. For finite $\tilde p_{\rm in}$ or finite $p\ri$ the proper time elapsed since $t_{\rm in}$ on a geodesic becomes infinite if $t_{\rm in}$ moves to the infinite past.

In contrast to our first model also the dimensionless proper time measured in units of inverse particle masses diverges. For $d\tilde \tau=\chi d\tau$ the relation between $\tilde\tau$ and $\gamma$ reads
\be\label{24N5}
\frac{d\tilde\tau}{d\gamma}=\chi\frac{d\tau}{d\gamma}=-
\frac{\chi}{\tilde H(\gamma^2-1)},
\ee
with 
\be\label{24N6}
\tilde H=\frac{\partial\ln (a\chi)}{\partial t}=H+\frac{\dot\chi}{\chi}=H-\frac{1}{3t}.
\ee
In the asymptotic limit $t\to-\infty$ one has $\tilde H\approx H$ such that 
\be\label{24N7}
\frac{\chi}{\tilde H}\approx \frac{\sqrt{3}m}{\mu}.
\ee
The solution of eq. \eqref{24N5} reads
\be\label{24N8}
\tilde\tau-\tilde\tau_{\rm in}=
\frac{\sqrt{3}m}{2\mu}\ln
\left(\frac{(\gamma+1)(\gamma_{\rm in}-1)}{(\gamma-1)(\gamma_{\rm in}+1)}\right).
\ee
For finite $\tilde p_{\rm in}$ or finite $p\ri$ the difference $\tilde \tau-\tilde \tau_{\rm in}$ diverges in the limit $t_{\rm in}\to-\infty$ since $\gamma(t)$ approaches one according to eq. \eqref{19n}. Again, $t_{\rm in}\to-\infty$ corresponds to the infinite past as measured with the ``clock'' $\tilde \tau$ of massive particles. For particles at rest one finds the explicit expression
\be\label{8n}
\tilde \tau=\int^\tau_0dt \chi(t)=-\frac{3m}{2}
\left(\frac{2t^2}{\sqrt{3}\alpha^2\mu}\right)^{\frac13}.
\ee

\subsection{Infinite physical time}
The trajectories of massless particles or photons are conveniently studied by switching to conformal time, $ds^2=a^2(\eta)(-d\eta^2+dx^kdx^k),~dt=a(t)d\eta$. We find an asymptotic behavior for $t\to-\infty$,
\be\label{9n}
\eta(t)=-\frac{1}{H(t)a(t)}~,~t(\eta)=-\frac{2\alpha}{\sqrt{3}\mu}\Big(\ln(-\mu\eta)\Big)^{3/2},
\ee
such that conformal time diverges in the infinite past, $\eta(t\to-\infty)\to-\infty$. Photons travel on straight lines in arbitrary directions, with $|dx|=d\eta$. The distance in comoving coordinates that they have moved from $t_{\rm in}$ to $t$ is given by 
$\Delta x(t,\ti)=\eta(t)-\eta(\ti)$. For a fixed $t$ it diverges for $\ti\to-\infty, \Delta(t,t_{\rm in}\to -\infty)\to \infty$. This is a similar qualitative behavior as for Minkowski space. 

For photons (or photon-like particles) proper time is no longer available for the definition of a coordinate-invariant physical time. As mentioned before, we may use instead the number of oscillations of the amplitude. For our coordinate system this ``oscillation time'' is proportional to conformal time $\eta$ in the infinite past. For oscillation time the distance to the infinite past is therefore divergent, as it should be. We conclude that the solution \eqref{1n}, \eqref{4n} is free of singularities from the infinite past to the infinite future. 

\subsection{Interpretation of geodesic incompleteness in de Sitter space}
In the Einstein frame our second model is given by eqs. \eqref{6G}, \eqref{6H} with $\tilde \alpha=0$. For small $B$ or large $\alpha$ primordial cosmology amounts again to an inflationary epoch. The spectral index depends only on $N$, $n=1-2/N$, and the tensor amplitude is very small, $r=8/(\alpha^2N^2)$, cf. sect. \ref{Eternal Universe and inflation}. The time coordinate of the Robertson-Walker metric can now be continued to the infinite past, $t\to-\infty$, where geometry approaches de Sitter space,
\be\label{66A}
H=\frac{\mu M}{\sqrt{3}m}~,~\sigma=-\frac M\alpha\ln (c_\sigma-\alpha^2 Ht).
\ee

In the Einstein frame the particle masses are constant such that $\tilde p$ and $p$ or $\tilde \tau$ and $\tau$ can be used equivalently. Massive particles move on time-like geodesics. We recall that $\tilde \tau$, evaluated on massive particle trajectories, is a frame invariant quantity. Since for the infinite past $\tilde \tau$ becomes infinite in the freeze frame for asymptotic massive particles, we expect that the proper time elapsed from the infinite past to some finite $t_0$ should diverge for asymptotic massive particles in the Einstein frame as well. This seems to contrast with arguments that de Sitter space has a past ``singularity'' or a ``beginning'', based on geodesic incompleteness \cite{HE,BV,BGV,AV}. We will explain this apparent discrepancy by a careful investigation of the appropriate limits. This will shed further light on the physical meaning of 
incomplete geodesics.

For de Sitter space the Robertson-Walker scale factor obeys $a=\exp (Ht)$, with constant $H>0$. (More precisely we investigate the geometry spanned by the Robertson-Walker metric in the range $-\infty<t<\infty$, as obtained from the field transformation from the freeze frame where $-\infty<t<\infty$. Formally, it is possible in other coordinates to continue de Sitter space beyond the surface $a=0$. In our setting, however, the geometry of the freeze frame is mapped precisely to the part of de Sitter space $a>0,~t>-\infty$.) We want to see if $t\to-\infty$ can be associated with a ``past-eternal'' universe. 

Let us consider trajectories of particles with constant non-zero mass in de Sitter space. They move on geodesics $(u_0=u(t_0)$ etc.),
\ba\label{23}
u&=&u_0\exp \big\{-2H(t-t_0)\big\},\nn\\
\gamma&=&{\coth}\big\{H(\tau-\tau_c)\big\}.
\ea
The condition $a^2u^2=\gamma^2-1$ relates $\tau$ and $t$,
\be\label{24}
t=2t_0+\frac{1}{H}\ln
\left[u_0\sinh\big\{H(\tau-\tau_c)\big\}\right].
\ee
The difference $\tau_0-\tau_c$ obeys
\be\label{25}
\sinh \big\{H(\tau_0-\tau_c)\big\}=
\frac{1}{\sqrt{\gamma^2_0-1}},
\ee
and we infer $\tau_c<\tau_0$, with $\tau_c\to-\infty$ for $\gamma_0\to 1$. 

For $\tau\to\tau_c$ both $\gamma$ and $u$ diverge and cosmic time $t$ goes to minus infinity. Except for $\gamma_0=1$ the difference $\tau_0-\tau_c$ remains finite. Geodesics ``stop'' at finite proper time $\tau_c$. This is called ``geodesic incompleteness towards the past''. At first sight it seems that for ``most geodesics'' the limit $t\to -\infty$ is only a finite distance in proper time away. This is usually interpreted as sign of a ``beginning'', despite the fact that all invariants built from powers of the curvature tensor and its covariant derivatives remain finite. The static trajectory with $\gamma_0=1$, for which $t\to-\infty$ is an infinite distance in proper time away, is somehow discarded as ``being of measure zero''. In contrast, we will argue that $\gamma_0=1$ is actually the only value for which one is allowed to use proper time for $t\to-\infty$ in a consistent way. Instead of being an exception, this is the value appropriate for asymptotic massive particles. As a consequence, the geodesics of asymptotic massive particles are complete also towards 
the infinite past. For all particles that have started in the infinite past with finite momentum the proper time elapsed when they arrive at $t_0$ is infinite.

Due to Hubble damping the (squared) momentum is always decreasing as time increases. For nonzero momentum $\gamma$ is larger than one, and $\partial\gamma/\partial t=\partial\ln \gamma/\partial\tau$ is always negative. Thus $\gamma$ and therefore momentum decreases with increasing $t$. Finite momentum is equivalent to a finite value of $\gamma$. We want to compute for finite $t_0$ the allowed range of $\gamma(t_0)$ under the condition that the value $\gamma(\ti\to-\infty)$ remains finite. For this purpose we impose the bound $\gamma(\ti)<\gamma_{\rm max}$ with $\gamma_{\rm max}$ an arbitrarily large but finite value. For fixed $\gamma_{\rm max}$ we take the limit $\ti\to-\infty$, reflecting the condition of finite momentum in the infinite past. We use the relation
\be\label{26}
\gamma^2(t_1)-1=\big(\gamma^2(t_2)-1\big)\exp \big\{-2H(t_1-t_2)\big\},
\ee
with $t_1=t_0$ and $t_2=\ti$, in order to establish the bound
\be\label{27}
\gamma^2(t_0)-1<(\gamma^2_{\rm max}-1)\exp 
\big\{-2H(t_0-\ti)\big\}.
\ee
For finite $\gamma_{\rm max}$ the r.h.s. goes to zero as $\ti\to-\infty$. One concludes that the condition of finite momentum results precisely in $\gamma_0=1$. More generally, for fixed $\gamma_{\rm max}$ and $\ti$ there remains only a restricted range of allowed values for $\gamma_0$ due to the damping between $\ti$ and $t_0$. This range depends on $\ti$. It shrinks to a single point $\gamma_0=1$ if $\ti\to -\infty$. This behavior constitutes a further example for the ``focus property'' of a differential equation that we have discussed in sect. \ref{Focus property of primordial cosmology}.

We can compute the proper time elapsed between $\ti$ and $t_0$ for $\gamma(\ti)=\gamma_{\rm max}$. This is the minimum of the proper time for all particles with $\gamma(\ti)\leq \gamma_{\rm max}$. From 
\ba\label{S8}
&&\tau_0-\tau_{\rm in}=t_0-\ti\\
&&+\frac1H\ln\left(\frac{1+\sqrt{1+(\gamma^2_{\rm max}-1)\exp \big\{-2H(t_0-\ti)\big\}}}{\gamma_{\rm max}+1}\right)\nn
\ea
one infers the limiting behavior for $\ti\to-\infty$,
\be\label{S9}
\tau_0-\tau_{\rm in}=t_0-\ti-\frac1H\ln
\left(\frac{\gamma_{\rm max}+1}{2}\right).
\ee
Thus $\tau_0-\tau_{\rm in}$ is smaller than $t_0-\ti$, but only by a finite amount as long as $\gamma_{\rm max}$ remains finite. We conclude that the proper time difference $\tau_0-\tau_{\rm in}$ diverges for $\ti\to-\infty$ for any massive particle with finite momentum in the infinite past. In other words, the infinite past $t\to -\infty$ is at infinite proper time distance for all massive particles with finite momentum.

The interpretation of the geodesics that are incomplete in the past for $\gamma_0>1$ becomes now rather simple. All these geodesics correspond to infinite momentum in the past. With
\ba\label{49AA}
p(t)=\frac{a(t_0)}{a(t)}p(t_0)=m\sqrt{\gamma_0^2 -1} \frac{a(t_0)}{a(t)}
\ea
all particles with $\gamma_0>1$ behave as photons, $p^2/m^2\rightarrow\infty$, in the infinite past and proper time cannot be used as a useful measure of time. Only for asymptotic massive particles proper time remains a useful measure of time. The corresponding trajectories have all $\gamma_0=1$ and the proper time elapsed since the infinite past is infinite, as it should be. The incompleteness of geodesics for $\gamma_0>1$ does not indicate a singularity in space-time or an incompleteness of cosmology. It merely reflects that these geodesics cannot be realized by asymptotic massive particles.  We may summarize two key points: (i) Not all timelike geodesics can be realized by asymptotic massive particles. (ii) Cosmologies that approach de Sitter space in the infinite past can be considered as regular. The infinite past in cosmic time occurs for the infinite past in proper time for all asymptotic massive  particles. All other trajectories become photon-like in the infinite past 
and proper time has to be 
replaced by a more appropriate concept as oscillation time. The oscillation time distance to the infinite past is infinite for all particles.

\section{Eternal Universe and inflation}
\label{Eternal Universe and inflation}

Cosmologies with de Sitter inflation (geometry approaching de Sitter space in the infinite past) can describe an eternal universe. Known examples are higher dimensional inflation \cite{SW} or the models of cosmon inflation \cite{CWUE,CWCI}. There exist many other possibilities for an eternal universe. Beyond the examples discussed here an asymptotic behavior $H=\eta/(t_c-t)$ approaches flat space in the infinite past. For our general class of models \eqref{1a}, \eqref{2a} with arbitrary $A,B$ we find indeed the asymptotic behavior for $t\to-\infty$
\be\label{42AA}
H=\frac{\eta}{t_c-t}~,~\eta=\frac{2}{2-A}
\left(\frac BA-1\right).
\ee
(The qualitative behavior $H\sim t^{-1}$ has been suggested in ref. \cite{Pi} for the model of ref. \cite{CWQ2}, i.e. eqs. \eqref{1a}, \eqref{2a} with $V$ replaced by $V_0\sim \chi^{4-A}$. There seems to be no quantitative agreement of eq. \eqref{42AA} with ref. \cite{Pi}, however. For a model of inflation we need to employ the full potential \eqref{2a} since for $V_0$ inflation would not end.) Varying $A$ and $B$ we find a family of inflation models. For suitable values of these parameters the spectrum of primordial scalar and tensor fluctuations is compatible with observation.

\subsection{Freeze frame}
In the freeze frame we start with the field equations derived from the effective action \eqref{1a}, \eqref{2a} for arbitrary $A$ and $B$, generalizing eqs. \eqref{3a}, \eqref{5a},

\be\label{PC4}
\ddot{s}+3H\dot s+2\dot s^2=
\frac{\mu^2 x(A+2x)}{B(1+x)^2},
\ee
with 
\be\label{PC5}
H=\sqrt{\frac{\mu^2}{3(1+x)}+\frac{B\dot s^2}{6}}-\dot s.
\ee
For the inflationary epoch we neglect matter and radiation such that primordial cosmology only involves the cosmon coupled to gravity. The field equations have an exact solution 
\be\label{PC7}
\chi=0~,~H=0~,~R_{\mu\nu\rho\sigma}=0,
\ee
which is, however, unstable. The early stages of cosmology are described by the vicinity of this solution. For $A<2$ we can employ $x\ll 1$. 

For the approach to the infinite past we discuss the cosmology of a simplified model with potential
\be\label{PC7A}
V_a=\lambda\chi^{4-A},
\ee
where $\lambda=\mu^2 m^{A-2}$ sets the overall mass scale. This model describes the early stages of inflation, while the end of inflation will need the full potential \eqref{2a}. The field equations simplify correspondingly,
\ba\label{PC8}
\ddot{s}+3H\dot{s}+2\dot s^2&=&\frac{\mu^2 A}{B}e^{(2-A)s},\nn\\
H^2+2H\dot s+\left(1-\frac{B}{6}\right)\dot s^2&=&\frac{\mu^3}{3}e^{(2-A)s}.\label{PC9}
\ea

For $A<2$ we make the ansatz $(t<t_c)$
\ba\label{PC10}
e^{(2-A)s}=\frac{v}{\mu^2(t_c-t)^2}~&,&~H=\frac{\eta}{t_c-t},\\
\dot s=\frac{2}{(2-A)(t_c-t)}~&,&~\ddot{s}=\frac{2}{(2-A)(t_c-t)^2}.\nn
\ea
This yields algebraic equations for the constants $v$ and $\eta$, with solution 
\be\label{PC11}
\eta=\frac{2}{2-A}
\left(\frac{B}{A}-1\right)~,~v=\frac{1}{(2-A)^2}
\left(12\frac{B^2}{A^2}-2B\right).
\ee
The positivity of $v$ requires $A^2<6B$. (The second solution with $\eta_-=-2/(2-A)$ has always negative $v$.) For the interesting particular case $B=A$ one finds $\eta=0$ and space is flat in lowest order! This case was discussed in ref. \cite{CWVG} and corresponds to the model of sect. \ref{Crossover model with flat space in the infinite past}.  For $B>A$ the primordial universe expands, while for $B<A$ it shrinks. In all cases it approaches flat space in the infinite past for $t\to-\infty$ with an inverse power law for $H$. 

For fixed $A$ and $B\to\infty$ both $\eta$ and $v$ diverge. This also happens for fixed $B$ and $A\to 2$. For $A=2$ the power law solution \eqref{PC10} is no longer valid. In this case one finds an exponential behavior of $s$ with constant $H$, as discussed in ref. \cite{CWUE}. For $A>2$ no power law solution \eqref{PC10} exists either. 

\subsection{Einstein frame}

Despite the large variety of different behaviors for primordial cosmology the observational consequences for all these models are similar.  This is best seen in the Einstein frame. Performing the Weyl scaling \eqref{6F} the effective action becomes a standard inflationary model with fixed (reduced) Planck mass $M$ and effective action \eqref{6G}, \eqref{6H}. The ``primordial potential'' \eqref{PC7A} is transformed to an exponentially decaying potential
\be\label{PC14}
V'_0(\sigma)=\frac{\mu^2M^4}{m^2}\exp 
\left(-\frac{\tilde\alpha\sigma}{M}\right)~,~\tilde \alpha=\frac{A}{\sqrt{B}}.
\ee
The ratio $\mu^2/m^2$ can be absorbed by a shift in $\sigma$, such that the behavior only depends on $\tilde \alpha$. For small enough $\tilde \alpha$ this cosmology is a type of power-law inflation. It has, however, no end since no value of $\sigma$ is singled out. This holds despite the appearance of $t_c$ in eq. \eqref{PC10} which naively suggests an end of inflation for $t\to t_c$. In the Einstein frame, however, this limit corresponds to the infinite future.

For a realistic model of inflation and a computation of primordial density fluctuations the crossover behavior of the potential is crucial and we have to consider the full potential \eqref{6H}
\ba\label{PC15}
V'&=&\frac{\mu^2M^4}{m^2}\exp 
\left(-\frac{\alpha\sigma}{M}\right)
\left[1+\exp \left(-\frac{(\alpha-\tilde\alpha)\sigma}{M}\right)\right]^{-1},\nn\\
\alpha&=&\frac{2}{\sqrt{B}}.
\ea
For large $\sigma$ the last bracket approaches one, such that the universe can make a transition to a late cosmology. We require a large value of $\alpha> 10$ in order to guarantee a small fraction of early dark energy. For the full potential the parameters $A,B$ and $\mu^2/m^2$ all play a role. They determine the end of the inflationary epoch and the spectrum of the primordial density fluctuations.

Let us investigate a possible range of ``slow roll inflation'' with ``slow roll parameters''
\be\label{PC16}
\epsilon =\frac{M^2}{2}
\left(\frac{\partial\ln V'}{\partial\sigma}\right)^2=\frac12
\left(\frac{\tilde\alpha}{1+x}+\frac{\alpha x}{1+x}\right)^2,
\ee
and
\ba\label{PC17}
\eta&=&2\epsilon+M^2\frac{\partial^2\ln V'}{\partial\sigma^2}\nn\\
&=&\frac{\tilde\alpha^2(1-x)-\alpha^2x(1-x)+4\tilde\alpha\alpha x}{(1+x)^2},
\ea
with 
\be\label{PC18}
x=\exp \left\{\frac{(\alpha-\tilde\alpha)\sigma}{M}\right\}.
\ee
For $\tilde\alpha=0$ one recovers the model of sect. \ref{Crossover model for de Sitter inflation}. The slow roll condition $\epsilon\ll 1$ is obeyed for small $\tilde \alpha$ in the region of small enough $x$. We may associate the end of inflation with $x_f=1/\alpha~(\epsilon\approx1/2)$. For $\alpha\gtrsim 10$, as required by the bounds on early dark energy, the inflationary epoch happens for small $x\ll 1$, where we can approximate
\ba\label{PC19}
\epsilon=\frac12(\tilde\alpha+\alpha x)^2~,~\eta=\tilde\alpha^2-\alpha^2x.
\ea

\subsection{Spectrum of primordial fluctuations}
In order to compute the spectral index $n$ and the tensor amplitude $r$ we need to evaluate $x$ at a time corresponding to $N$ e-foldings before the end of inflation, when fluctuations of observable scales left the horizon, $N\approx 50-65$. With 
\ba\label{PC20}
N&=&\int Hdt=-\int 
\left(\frac{\partial \ln V'}{\partial \sigma}\right)^{-1}\frac{d\sigma}{M}=\int \frac{1}{\sqrt{2\epsilon}}
\frac{d\sigma}{M}\nn\\
&=&\frac{1}{\alpha-\tilde\alpha}\int^{x_f}_{x(N)}
\frac{dx}{\tilde\alpha x+\alpha x^2}
\ea
one finds
\be\label{PC21}
x(N)\approx\frac{\tilde\alpha}{\alpha}
\Big[\exp \big\{\tilde\alpha(\alpha-\tilde\alpha)N\big\}-1\Big]^{-1}.
\ee
The standard relations for the spectral index $n$ and the tensor to scalar ratio $r$ read 
\ba\label{PC22}
n&=&1-6\epsilon+2\eta\approx 1-2\alpha^2x(N)-\tilde \alpha^2,\nn\\
r&=&16\epsilon=8\big (\tilde\alpha+\alpha x(N)\big)^2.
\ea

Depending on the values of $\tilde \alpha$ and $\alpha$ one finds two regimes. For $\tilde \alpha\alpha\ll 1/N$ one can expand the exponential in eq. \eqref{PC21}, leading to 
\be\label{PC23}
x(N)=\frac{1}{\alpha^2N}~,~n=1-\frac2N~,~r=\frac{8}{\alpha^2N^2}.
\ee
These are the values found for $A=0$ in ref. \cite{CWSF}, corresponding to the model of sect. \ref{Crossover model for de Sitter inflation}. On the other hand, for $\tilde \alpha\alpha\gg 1/N$ one finds 
\be\label{PC24}
x(N)\ll\frac{\tilde\alpha}{\alpha}~,~n=1-\tilde\alpha^2~,~r=8\tilde\alpha^2.
\ee
This corresponds to our first model in sect. \ref{Crossover model with flat space in the infinite past}. Between the two limits one finds a smooth transition. 

\begin{figure}[h!tb]
\centering
\includegraphics[scale=0.5]{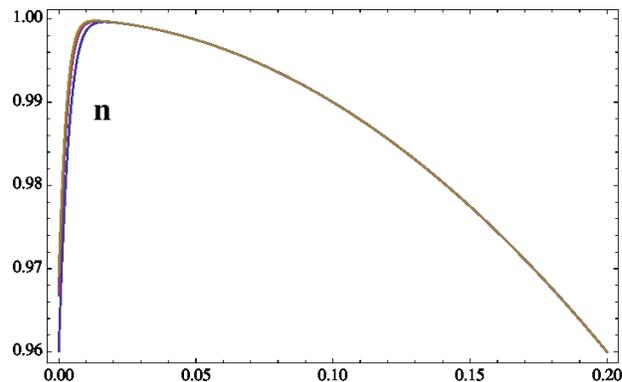}
\caption{Spectral index as a function of the parameter $\tilde \alpha$. We show curves for three different values $N=50,~60,~65$.}
\label{ETU2}
\end{figure}

We plot in Fig. \ref{ETU2} the spectral index as a function of $\tilde\alpha$, for $N=50,~60,~65$. Except for a range of very small $\tilde \alpha$ the spectral index is independent of $N$. Two regions of very small $\tilde\alpha$ or $\tilde\alpha$ near $0.2$ are consistent with Planck results \cite{Planck}. For the tensor ratio we may neglect the small contribution from $\alpha x\neq 0$ and use for practical purposes for all $\tilde \alpha$
\be\label{94}
r=8\tilde\alpha^2.
\ee
We observe that the allowed region of intermediate $\tilde\alpha$ could be consistent with the claimed detection of tensor modes by BICEP \cite{BICEP}. In this region one has the relation 
\be\label{95}
n=1-\frac r8.
\ee

The observed amplitude of the primordial fluctuations fixes the ratio $\mu/m$.

\section{Conclusions}
\label{Conclusions}

In this note we have put the emphasis on ``past eternity''. ``Future eternity'' for $t\to+\infty$ is rather generic for many cosmologies, including the Friedman universe. For our models \eqref{1a}, \eqref{2a} the universe produces entropy after the end of inflation \cite{CWCI,HMSS}. The subsequent radiation- and matter-dominated periods correspond to the approach to a fixed point for $\chi\to\infty$ for which scale symmetry becomes an exact symmetry which is spontaneously broken \cite{CWQ2,CWVG}. For $\chi^2\gg m^2$ we assume that the masses of all particles except for neutrinos scale $\sim \chi$ (perhaps with different coefficients as for $\chi^2\ll m^2$), and dimensionless gauge and Yukawa couplings are close to their constant fixed point values. Bounds on the time variation of fundamental constants are therefore obeyed. During radiation- and matter-domination the universe shrinks in the freeze picture, with slowly increasing temperature and particle masses \cite{CWUE,CWSF}. The scaling solution predicts a small fraction of early dark energy \cite{CWQ2,CWA,DLW,SDW,CWE,DR}, $\Omega_e=n/\alpha^2$, with $n=4(3)$ for radiation (matter) domination. We may assume that for the present epoch the neutrino masses increase faster than $\sim \chi$ (in the freeze frame), due to a crossover in a sector of heavy singlets \cite{CWVG}. Neutrinos becoming non-relativistic trigger a recent transition to a dark energy dominated epoch, with present dark energy related to the average neutrino mass \cite{GNQ1,GNQ2}. The models are compatible with all present cosmological observations \cite{CWSF,CWUE}. A 
measurement of primordial tensor fluctuations \cite{BICEP} will restrict the allowed ranges of $\tilde \alpha$ and $\alpha$.

In conclusion, we have presented consistent cosmological models for which solutions of field equations can describe an eternal universe, in contrast to the opinion based on earlier ``no-go theorems''. This does not imply that the history of the universe must have followed these solutions since the infinite past. Since the solutions are stable attractors, many other possibilities for a primordial universe can approach such attractors as time increases. Information on the primordial state is then largely lost - predictions for observations will be the same as for a primordial state following the ``eternal attractor solution'' since the infinite past. One could imagine a chaotic inflation \cite{Lin,SW} primordial state, governed by quantum fluctuations in flat space. Once a region is homogeneous enough such that the homogeneous field equations become valid, it will subsequently follow the inflation history according to the eternal attractor solution. 

Our approach allows for a differentiated view of several basic cosmological concepts. No big bang singularity is needed. The big bang picture in the Einstein frame provides for a very useful description of observations, but may be inappropriate for a good picture of the regular structure of the eternal universe. Gravity needs not to become strong in the ``beginning'' of the universe. The concept of the quantum effective action assumes a quantum field theory for gravity. Nevertheless, for our solutions gravity remains always a weak interaction. The concepts of time and geometry are ambiguous. This extends beyond the issue of general coordinate transformations. Field transformations leave observables invariant, but can map very different geometries into each other. Apparent singularities in a given frame may be ``field singularities'' associated to a particular choice of fields, while physical observables remain regular. 

\bigskip\noindent
{\bf Acknowledgment.} The author would like to thank P.~Steinhardt for stimulating discussions that motivated this work, and A.~Linde and A.~Vilenkin for useful comments.

\vspace{2.0cm}\noindent

\bibliography{eternal_universe}

\end{document}